# A plasmoid-model for mass loss from stars on the upper red giant branch: The mass loss rate is controlled by the number of density scale heights in the convection zone

J. MacDonald and D. J. Mullan

Dept of Physics and Astronomy, University of Delaware, Newark DE 19716. U.S.A.

**Abstract**

Recent asteroseismic determinations of $\Delta M$, the integrated mass loss on the red giant branch (RGB), for fields stars show a trend of $\Delta M$ *decreasing* as metallicity *increases.* This trend among field stars is inconsistent with many existing models of RGB mass loss. The present paper is motivated by a 'plasmoid' model of RGB mass loss in which magnetic flux loops, generated by a shear dynamo operating below the convection zone, are buoyed up to the stellar surface starting at the evolutionary stage right after the RGB 'kink'. This model leads us to examine correlations between, on the one hand, the average post-kink RGB mass loss rate, determined from $\Delta M$ and the post-kink RGB lifetime, and on the other hand, stellar properties which exist just after the end of the kink. For three distinct stellar samples, we find strong *anti-correlations* between the average post-kink RGB mass loss rate and the number of density scale heights in the convection zone. This leads us to propose that the number of density scale heights in the convection zone is a dominant factor in determining the rate of the mass loss process which sets in after the RGB kink.

## 1. Introduction: Mass loss rates as a function of stellar parameters

The Sun undergoes a continual process of mass loss. Parker (1958) proposed the earliest solar wind model in terms of the hydrodynamic expansion of a hot corona: this model predicted that the wind speeds would reach values of several hundreds of km s$^{-1}$ by the time the wind reached the Earth's orbit. Quantitative support for Parker's predictions of speed, density, and temperature have been obtained by many space-borne instruments over the past 60+ years (e.g. Mariner II: Neugebauer & Conway 1962; Parker Solar Probe: Kasper et al. 2019). In its current evolutionary state, a solar-like star loses mass at a rate which is so small that during its main sequence lifetime, the integrated amount of mass lost by the star is orders of magnitude smaller than the total mass of the star. Therefore, even if we did not have a reliable model of the solar mass loss process, it would still be possible to follow solar evolution fairly realistically for most of its lifetime.

On the other hand, many massive stars lose a very significant fraction (several tenths or more) of their mass in the course of their evolution. The most extreme example of integrated mass loss occurs in certain supernovae which eventually leave no stellar remnant at all after an explosive end-point (e.g. Renzo & Smith 2024). In such conditions, attempts to calculate the evolution of massive stars would be subject to significant errors if the mass loss process were not incorporated by means of a reliable model. Fortunately, such a model exists, based on the theory of radiative line driving in an expanding wind around a hot luminous star (Castor et al. 1975).

Here, we are interested in intermediate cases of integrated mass loss which occur among low-mass stars at certain phases of their evolution. In particular, our interest is in the integrated amount of mass loss $\Delta M$ which occurs during the red giant branch (RGB). More specifically, and guided by availability of appropriate data, we focus on the interval of time between two "markers" which exist on the evolutionary tracks of low-mass stars: the "kink" and the "red clump". The "kink" occurs at an evolutionary stage when a star is *ascending* the RGB to higher luminosities, but *before* the star has reached the tip of the RGB. The "red clump" occurs *after* the star has evolved through the tip of the RGB and then *descended* to the lower luminosities which are more characteristic of the horizontal branch (HB).

Access to **asteroseismic** data from the Kepler and TESS missions has opened up a window to evaluate more precise structural parameters in a large sample of stars. Of particular interest to us here is the ability to determine stellar masses with enough reliability that the *difference* in masses between the two evolutionary "markers" mentioned above can be determined with a precision of 10% or better. E.g., for the globular cluster M4, photometric data from the K2 mission has been used to determine that the integrated mass loss between RGB and HB stars is $\Delta M = 0.17 \pm 0.01$ M$_\odot$ (Howell et al 2022) or $\Delta M = 0.227 \pm 0.028$ M$_\odot$ (Tailo et al. 2022). And using Kepler data on 99 field stars with metallicities ranging from [Fe/H] = -0.9 to [Fe/H] = -0.47 on the RGB and 110 field stars with the same range in metallicity near the red clump, Brogaard et al. (2024) found $\Delta M = 0.167 \pm 0.007$ M$_\odot$.

Remarkably, although the integrated $\Delta M$ is now being determined to a precision approaching 3 significant places, nevertheless, the physical process of mass loss as a function of stellar parameters in red giants is still not reliably identified. There has been no shortage of theories: 25 years ago, Catelan (2000) was already able to list six different analytical formulas which had been proposed to describe the mass loss rate as a function of stellar parameters. The most widespread formula for mass loss rate from red giants is that of Reimers (1975): $dM/dt = -\eta L_* / g_* R_*$, where * denotes the stellar values in units of the respective solar values, and $\eta$ is a proportionality constant. In effect, this formula assumes that a constant fraction of the stellar output power is converted (by an unspecified process) to mass loss. The various formulas discussed by Catelan (2000) are based on different arguments, but despite this, most of the mass loss formulae can obtain a fairly good fit (within sometimes large error bars) to certain samples of stellar data. As a result, Catelan (2000) concludes with a remarkably firm statement as regards the need for incorporating mass loss into evolution codes: "*In effect, mass loss on the RGB is an excellent, but virtually untested, second-parameter candidate*" (italics in original). Subsequently, Catelan (2009) noted that the Reimers formula "does not properly account for the mass loss rates that have become available in the more recent literature": moreover, if the constant $\eta$ is set too high, the star will never reach the RGB tip, but will instead evolve at constant $L$ to implausibly high $T_{eff}$ (Cassisi & Salaris 1997). Furthermore, Catelan (2009) also reported that putative dependences of the various mass loss formulae on metallicity results in metal-poor stars having systematically smaller $\Delta M$.

An independent comparison between mass loss formulae and observational data in the infrared (Origlia et al. 2002) suggests that the mass loss rates might *not* depend clearly on *L, g,* or *R*. Moreover, the mass loss might be episodic, with each episode lasting longer than a few days, and occurring during the evolutionary phase close to the RGB tip. However, Boyer et al. (2008) have expressed some caveats about the data analysis of Origlia et al. due to blends in their infrared images.

Cranmer & Saar (2011: CS11) developed a model for *dM/dt* in cool stars based on the theory of MHD turbulence in the presence of thermally-driven convection. The theory requires specification of various parameters: density scale height, equipartition field strength, the Rossby number, Alfven wave fluxes $F_{A*}$ in the photosphere, filling factor of the magnetic field in the photosphere $f_*$ and also in the transition region, the length scale of the largest turbulent eddies, and the conductive flux. CS11 made a number of approximations in order to assign plausible values to these parameters. Two limiting cases were considered: (i) Coronal: the MHD dissipation occurs in rarefied gas where coronal temperatures ($T \approx 10^6$ K) develop; (ii) Chromospheric: the gas is so dense that radiative losses limit the temperatures to chromospheric values ($T < 10^4$ K). When compared with observations of 47 cool stars, various choices of parameters lead to reduced $\chi^2$ values <1 indicating statistically significant fits: however, the spread in individual mass loss rates at any given luminosity may be as large as 3-4 orders of magnitude. For main sequence stars, CS11 derive a simplified formula (their eq. 45) for *dM/dt* in terms of $R_*$, $L_*$, $F_{A*}$, and $f_*$. However, CS11 caution that their eq. 45 applies only to main sequence stars, and therefore should not be applied to the evolved stars which are of interest to us here.

In regard to the dependence of integrated mass loss on metallicity, Li (2025) has reported on a mixed sample of stars which includes some stars from open clusters and some stars from globular clusters. For this mixed sample, Li (2025) finds that as a function of [Fe/H], the integrated $\Delta M$ *increases* as [Fe/H] decreases from +0.4 to roughly -0.5. But as [Fe/H] decreases further from -0.5 to -2.5, the opposite behavior occurs: the integrated $\Delta M$ *decreases.* It seems that when we examine a mixed sample of stars from different environments, the effects of metallicity on mass loss rates may not be monotonic.

In the present paper, our goal is to determine if the observed values of integrated mass loss $\Delta M$ can be understood in terms of a model in which mass loss starts to operate at a specific point ("the kink") in the evolutionary track of a star ascending the RGB.

## 2. Observational evidence for onset of significant mass loss in cool giants

When the International Ultraviolet Explorer (IUE) satellite was launched in 1978 to study the ultraviolet spectra of stars between wavelengths of ~1200 Å and ~3000 Å, one of its spectrographs (with resolving power ~$10^4$) enabled the study of profiles of the *h* and *k* lines of singly ionized magnesium (at wavelengths of 2806 and 2796 Å) with a wavelength resolution of about 0.3 Å. At this resolution, Stencel & Mullan (1980: SM80) discovered that, in a sample of 47 cool giant stars, the *h* and *k* lines exhibit emission reversals at the center of a broad absorption profile. Moreover, the emission reversals of both *h* and *k* lines could be clearly resolved into two peaks, one shortward of line center (*S*), and one longward (*L*). The *S* and *L* components were separated by up to 1 Å or more When results for individual cool giants were plotted in the color-magnitude diagram (CMD), SM80 noted that stars lying *below and to the left* of a certain

"dividing line" in the CMD essentially all displayed an *S* peak stronger than the *L* peak, i.e. *S/L* > 1. In contrast, stars lying *above and to the right* of the same "dividing line" essentially all displayed an *L* peak stronger than the *S* peak, i.e. *S/L* < 1. SM80 suggested that a cool wind flowing out from the star could provide a natural explanation for the behavior *S/L* < 1: such a wind, expanding along the line of sight *towards the observer* could give rise to *blue-shifted* absorption. If the wind moves at speeds of several tens of km s$^{-1}$, the blue-shifts amount to a few times 0.1 Å: this would enable the wind to absorb *S* peak photons more effectively than absorbing *L* peak photons, leading to *S/L* < 1. In view of this, SM80 proposed a "velocity dividing line" (VDL) in the CMD as a marker of the onset of a wind of considerable absorbing power in cool giants. However, a quantitative estimate of the mass loss rate in such a wind would have required a sophisticated NLTE analysis of the atomic energy levels of MgII: SM80 made no attempt to undertake such an analysis. Their results were confined primarily to identifying a *location* in the CMD were mass loss would switch from (effectively) "off" to (effectively) "on" among cool giants.

A separate "temperature dividing line" (TDL) had already been reported in IUE data (Linsky & Haisch 1979) separating stars with hot coronae from stars with cool chromospheres. SM80 showed that there was a fairly close overlap of the two dividing lines, indicating that mass loss in cool giants was associated predominantly with cool winds. On the other hand, "coronal stars", with *S/L* generally > 1 showed no evidence for detectable cool wind outflows: such stars might have hot winds which are too highly ionized to absorb the *S* peak of singly ionized magnesium.

### 3. Association of mass loss onset with the RGB "kink"

The possibility that the onset of mass loss in cool giants might be associated with an evolutionary "event" while a cool star is ascending the RGB was suggested by Mullan & MacDonald (2003: MM03a). The event in question occurs when the growing He core pushes an H-burning shell outward until the shell encounters a region where the local H has never previously been exposed to nuclear burning. At this point, the H-burning shell suddenly has access to an enhancement in fuel for H-burning, resulting in a sudden but temporary reduction in surface luminosity, after which the star continues its ascent up the RGB. [This behavior had been predicted on theoretical grounds by Thomas (1967) and by Iben (1968) before observers reported it.] The result is a "kink" in the evolutionary track, in which the rate of evolution is temporarily slowed, thereby leading to a localized enhancement in the number of stars in that region of the CMD. This localized enhancement in numbers of stars appears as a "bump" in the luminosity function of globular clusters (see Cassisi & Salaris 1997, and references therein).

When MM03a plotted data for several cool giants in the CMD, the data suggested that stars displaying *S/L* <1 tended to lie above and to the right of the "kink", while stars with *S/L*>1 tended to lie below and to the left of the "kink". Based on this result, MM03a suggested that the onset of significant mass loss in cool winds from cool giants might be associated with evolution through the "kink".

In this regard, it is worth noting that, in an asteroseismic study of stars in the globular cluster M4, Howell et al. (2022) report that there appears to be a "down-step" in the mean mass between stars on the lower RGB and stars on the upper RGB. The "down-step", amounting to 0.03 - 0.17 $M_\odot$ (depending on which asteroseismic formula one uses), is observed to occur at a $G_{mag}$ which coincides with the location of the "kink".

## 4. Rotationally driven shear instability in a star at the "kink"

In a follow-up paper to MM03a, MacDonald & Mullan (2003: MM03b) presented a theoretical analysis of the onset of shear instability inside a rotating star on the RGB. When a star evolves up through the "kink", a region of shear instability emerges just beneath the lower boundary of the convective envelope. MM03b argue that the instability gives rise to a large number of small magnetic loops which are driven upward towards the surface by buoyancy forces. Some of the loops will be dissipated in the convection zone, by diffusion and/or reconnection. But a finite number of loops could survive to the stellar surface, and MM03b postulated that these could provide a mechanism to drive mass loss from the star. Such a process would be distinct from the magnetically-driven wind described by CS11: in the latter, the magnetic fields originate in thermally-driven convective instability *within* the stellar envelope. But in our case, the fields originate in a region of rotationally-driven shear instability *beneath* the convective envelope.

The buoyantly rising loops will not only cause mass loss to occur. Other effects would also accompany the rising loops: specifically, enhanced abundance of the $^{13}C$ isotope, and inhibition of lithium depletion.

In 2003, it was difficult to develop a quantitative model of the MM03b postulate: there were too many unknowns. However, by 2019, more information had become available, and this enabled us to examine the postulate by means of a quantitative model. We now turn to a summary of that model.

## 5. The concept of a plasmoid-driven wind

Mullan & MacDonald (2019: MM19) suggested that, in order to deal with the emergence of many magnetic loops at the surface of a star, it could be worthwhile to consider a "plasmoid model" which had been proposed originally for the solar wind by Pneuman (1983). Using available data on stellar properties, MM19 derived an estimate of the number of magnetic loops which might survive the traverse from the shear unstable layer below the convective envelope to the stellar surface. Many of the loops, after broaching the surface, could be expected to expand upwards into the lower density atmosphere, and reconnection might eventually "pinch off" the feet points of the loops. In this way, a supply of plasmoids would be launched as a wind from the star. Available information on number densities of the atmosphere, combined with estimates of the number of loops on the surface, enabled an estimate of the mass loss rate. The nominal result was predicted to be of order $10^{-8} M_\odot$ yr$^{-1}$. In the various samples of stars which we examine below, it will be important to determine whether or not this nominal prediction of the mass loss rate is consistent with empirical data.

Following Pneuman's theory, the wind is expected to accelerate more or less rapidly away from the star depending on what fraction of the wind mass was contributed by plasmoids.

Variations in the supply of plasmoids from one star to another are expected to lead to the possibility that the above nominal result might be uncertain by as much as an order of magnitude on either side of the nominal mass loss rate.

## 6. Results from evolutionary models

### 6.1. Dependence of the average post-kink mass rates on stellar properties at the kink

In order to determine how integrated mass loss and average mass loss rates correlate with properties of stellar models at the location of the kink on the RGB, we have used the DEUCES code (Lawlor & MacDonald 2023) to evolve models of stars of initial masses in the range $M_0 = 0.7 – 1.20\,M_\odot$, from the pre-main sequence phase to the onset of the helium core flash without mass loss. The ranges of [Fe/H], [α/Fe] and helium mass fraction $Y$ are chosen to cover those of stars discussed by Tailo et al. (2020), Tailo et al. (2025) and Li (2025). For simplicity, convective overshoot is not included.

For models of low stellar mass and low values of [Fe/H] there is no well-defined kink in the HRD on the RGB when the nuclear burning front reaches the hydrogen discontinuity. However there still is a slowdown in evolution at this point which would lead to a luminosity function bump. We, therefore, determine $\tau_{postkink}$ by taking the time of the kink to be when the time derivative of the luminosity is minimum.

Once the values of $\Delta M$, the observationally determined integrated mass loss on the RGB are known, we use our models to divide $\Delta M$ by the time taken to evolve from the kink to the core flash: this provides us with an estimate of the average post-kink mass loss rate, $\bar{\dot{M}}$. We have compared various properties of the models at the kink with $\bar{\dot{M}}$, and find, for certain data sets described below, a strong correlation between the number of density scale heights in the red giant model convection zone at the evolutionary point when the stellar luminosity begins to increase again just after the kink with $\bar{\dot{M}}$.

### 6.2. Results for field and open cluster stars

We have used results from Li (2025) to determine the trend of $M_0$, $\Delta M$, and [α/Fe] with [Fe/H] for field and open cluster stars. We determine the helium mass fraction, $Y$, by converting [Fe/H] and [α/Fe] to a heavy element mass fraction $Z$ and assuming that $Y$ increases linearly with $Z$. The zero point is set by the big bang nucleosynthesis value, $Y_0 = 0.2484$ (Cyburt et al. 2003). In principle, the slope can be determined from estimates of proto-solar abundances. There have been a number of determinations of proto-solar abundances over the years Values of $dY/dZ$ range from 0.97 (Anders & Grevesse 1989; Gruberbauer 2012) to 1.72 (Lodders 2003). We opted for a value near the middle of the range, and our specific relation is

$$Y = 0.2485 + 1.31Z. \qquad (1)$$

Our adopted values for the parameters together with the time for the model to evolve from the kink to the core helium flash, $\tau_{postkink}$, the average post-kink mass loss rate, $\bar{M}$, and the number of density scale heights in the convection zone just after the kink, $n_{dsh}$, are given in Table 1. The masses given in the 4$^{th}$ column of Table 1 show the range in mass that we used in our modeling and are based on values given by Li (2025) for stars with [Fe/H] and [α/Fe] values similar to those given in the table. The uncertainties given for $\tau_{postkink}$ and $n_{dsh}$ result directly from the mass uncertainty. The values for $\Delta M$ and its uncertainty are determined from the linear regression performed by Li. The uncertainties in $\tau_{postkink}$ and $\Delta M$ are then propagated to find the uncertainty in $\bar{M}$. For this sample, the average mass loss rates are less than the nominal rate mentioned in Section 5 by factors of 2 or more.

**Table 1**
Data and model parameters for field and open cluster stars

| [Fe/H] | [α/Fe] | Y | $M_0$ (M$_\odot$) | $\Delta M$ (M$_\odot$) | $\tau_{postkink}$ (10$^6$ yr) | $\bar{M}$ (10$^{-8}$ M$_\odot$ yr$^{-1}$) | $n_{dsh}$ |
|---|---|---|---|---|---|---|---|
| -0.80 | 0.3 | 0.255 | 0.95-1.00 | 0.204±0.013 | 47.80±1.34 | 0.428±0.039 | 13.186±0.020 |
| -0.65 | 0.25 | 0.256 | 0.95-1.00 | 0.177±0.012 | 53.85±1.76 | 0.330±0.033 | 13.400±0.021 |
| -0.50 | 0.2 | 0.258 | 0.95-1.00 | 0.150±0.012 | 60.64±1.80 | 0.248±0.027 | 13.627±0.021 |
| -0.25 | 0.1 | 0.263 | 1.05-1.15 | 0.105±0.010 | 62.22±2.25 | 0.170±0.022 | 13.896±0.042 |
| 0.00 | 0.0 | 0.270 | 1.10-1.20 | 0.060±0.010 | 71.43±4.07 | 0.085±0.019 | 14.208±0.040 |
| 0.20 | 0.0 | 0.282 | 1.10-1.20 | 0.024±0.010 | 79.62±4.52 | 0.031±0.014 | 14.461±0.039 |
| 0.35 | 0.0 | 0.294 | 1.10-1.20 | 0.02±0.02 | 83.44±4.85 | 0.025±0.025 | 14.609±0.039 |

The correlation between $\bar{M}$ and $n_{dsh}$ can be clearly seen in Figure 1.

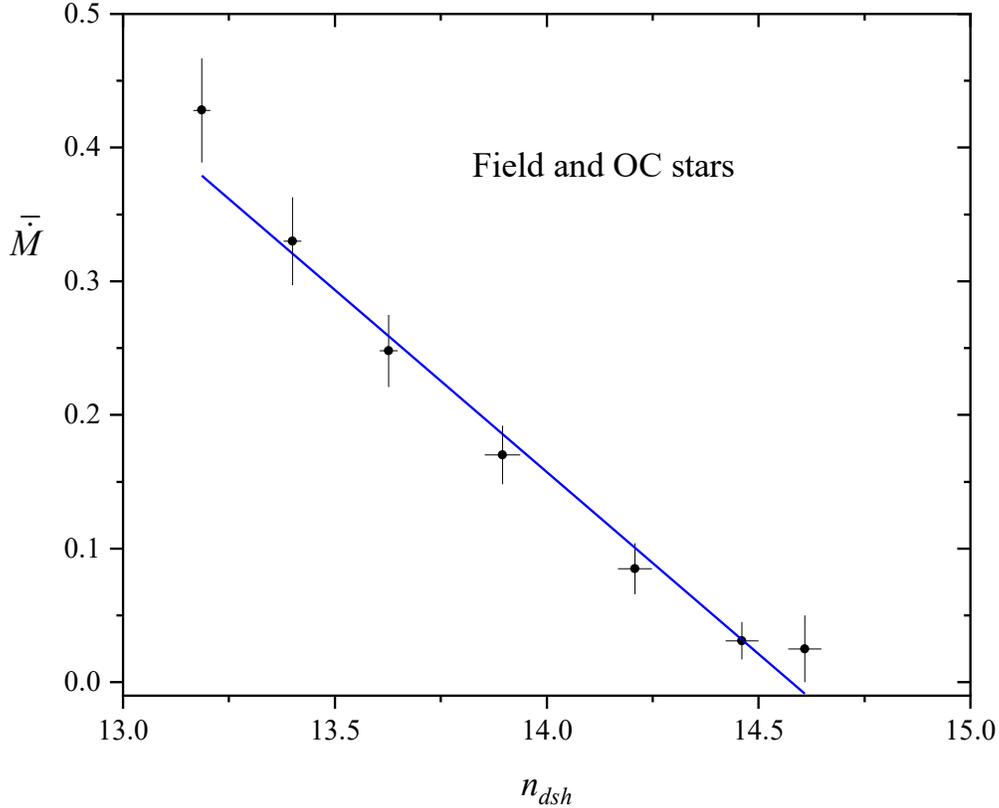

Figure 1. The average post-kink RGB mass loss rate in units of $10^{-8}$ M$_\odot$ yr$^{-1}$ for field and open cluster stars is shown as a function of the number of density scale heights in the convection zone just after the RGB kink. The blue line shows the best linear regression fit to the data.

Linear regression applied to our results shown in fig. 1 gives

$$\frac{\overline{\dot{M}}}{10^{-8}\ \mathrm{M}_\odot\ \mathrm{yr}^{-1}} = (3.97 \pm 0.28) - (0.273 \pm 0.020)\, n_{dsh}, \quad (2)$$

with an adjusted R-square value of 0.969. This strong (anti)correlation is highly suggestive that the mass loss rate may be controlled by the depth in scale heights of the red giant's convection zone, as might be expected for the plasmoid driven wind discussed in section 5.

### 6.3. Results for first generation stars in globular clusters

Multiband HST photometry (Piotto et al. 2015) has revealed that nearly all globular clusters have multiple populations (Milone et al., 2015a, b; Nardiello et al., 2015; Tailo et al. 2020). A distinguishing feature of individual populations is the anticorrelation between O and Na abundances (Gratton, Sneden & Carretta 2004), which is attributed to high temperature proton capture reactions that occurred in earlier generations. These reactions lead to depletion of C and O and enhancement of He, N and Na (Denissenkov & Denissenkova 1990; Langer, Hoffman &

Zaidins 1997). The abundance patterns observed in GCs allows separation of stars into first generation (1G) and later generations, usually collectively referred to as second generation (2G).

**Table 2**
Data and model parameters for first generation globular cluster stars

| [Fe/H] | [α/Fe] | Y | $M_0$ ($M_\odot$) | $\Delta M$ ($M_\odot$) | $\tau_{postkink}$ ($10^6$ yr) | $\bar{\dot{M}}$ ($10^{-8}$ $M_\odot$ yr$^{-1}$) | $n_{dsh}$ |
|---|---|---|---|---|---|---|---|
| -2.4 | 0.4 | 0.25 | 0.78-0.79 | 0.085±0.020 | 21.56±0.11 | 0.395±0.095 | 11.512±0.002 |
| -2.2 | 0.4 | 0.25 | 0.78-0.79 | 0.104±0.019 | 23.79±0.26 | 0.438±0.085 | 11.669±0.002 |
| -2.0 | 0.4 | 0.25 | 0.80-0.81 | 0.123±0.019 | 25.36±0.12 | 0.485±0.077 | 11.823±0.001 |
| -1.75 | 0.4 | 0.25 | 0.79-0.80 | 0.147±0.017 | 30.33±0.20 | 0.485±0.059 | 12.093±0.001 |
| -1.5 | 0.4 | 0.25 | 0.81-0.82 | 0.171±0.016 | 34.80±0.22 | 0.492±0.049 | 12.369±0.002 |
| -1.25 | 0.4 | 0.25 | 0.82-0.83 | 0.194±0.015 | 41.67±0.30 | 0.466±0.039 | 12.691±0.003 |
| -1.0 | 0.4 | 0.25 | 0.83-0.86 | 0.219±0.014 | 51.21±1.00 | 0.428±0.036 | 13.082±0.010 |
| -0.75 | 0.3 | 0.25 | 0.87-0.91 | 0.242±0.013 | 57.74±1.25 | 0.420±0.032 | 13.369±0.015 |
| -0.5 | 0.2 | 0.25 | 0.90-0.92 | 0.266±0.012 | 68.95±0.82 | 0.386±0.022 | 13.724±0.008 |

In this section, we investigate whether a correlation between $\bar{\dot{M}}$ and $n_{dsh}$ does or does not exist for the first-generation stars in globular clusters by using the results for RGB mass loss from 46 globular clusters presented in Tailo et al. (2020). The data used is given in Table 2 together with the model derived parameters, $\tau_{postkink}$, $\bar{\dot{M}}$, and $n_{dsh}$. The masses given in the 4th column of Table 2 show the range in mass that we used in our modeling and are based on values given by Tailo et al. (2020) for globular clusters with [Fe/H] and [α/Fe] values similar to those given in the table. Uncertainties are determined in the same way as described in the previous subsection but using data from Tailo et al. The values for $\Delta M$ and its uncertainty are determined from the linear regression performed by Tailo et al.

Figure 2 shows that statistically there is a trend of average post-kink mass loss rate increasing as $n_{dsh}$ decreases. Linear regression applied to our results shown in fig. 2 gives

$$\frac{\bar{\dot{M}}}{10^{-8} \text{ M}_\odot \text{ yr}^{-1}} = (1.046 \pm 0.168) - (0.0474 \pm 0.0128) n_{dsh}, \quad (3)$$

with adjusted R-square value 0.613. This represents a weaker (anti)correlation than was found above for field stars and stars in open clusters. Even so, the slope is found to be significantly different from zero. Fig.2 suggests that the increase of average mass loss rate with decreasing $n_{dsh}$ might not hold for [Fe/H] less than about -2.0. If the two lowest [Fe/H] values are removed from the sample, the best fit linear regression line (shown in fig. 2 by the red line) is found to have an adjusted R-square value of 0.939, comparable to that found for the field and open cluster stars.

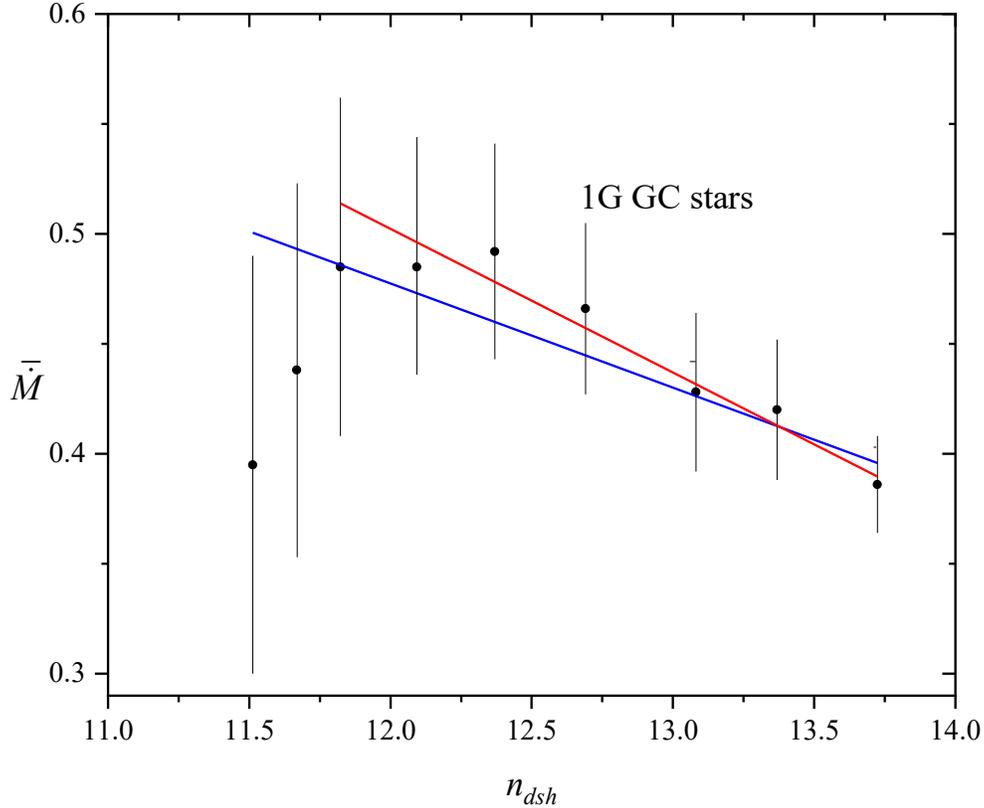

Figure 2. As fig. 1 but for first generation stars in globular clusters. The blue line shows the best linear regression fit to the data. The red line shows the best fit when the two data points corresponding to the lowest [Fe/H] values are omitted.

For this sample, the average mass loss rates differ from the nominal rate mentioned in Section 5 by factors of 2.5 or less.

*6.4. Results for stars in the multiple populations of globular clusters NGC 6752 and NGC 2808*

Tailo et al. (2025) have investigated how the integrated mass loss on the RGB varies amongst the populations of the GCs NGC 6752 and NGC 2808. They conclude that the integrated mass loss $\Delta M$ is higher in later generations than in the first generation. They also conclude that there is also a correlation between $\Delta M$ and the helium enhancement of the population. NGC 6752 has three populations (Milone et al. 2013) and NGC 2808 has five populations (Milone et al. 2015) but spectroscopy is only available for four of these populations (Carretta et al. 2006). We give in Table 3 the data from Tailo et al. (2025) used in our analysis together with the model derived parameters. Since the O abundance varies between populations, there is uncertainty in which [α/Fe] to use in the stellar modelling. We base our choice of [α/Fe] on the [Mg/Fe], [Si/Fe] and [Ca/Fe] ratios determined by Milone et al. (2013) for all three NGC 6752 populations and the [Mg/Fe] ratio determined by Milone et al. (2015) for two of the NGC 2808 populations.

Table 3
Data and model parameters for the globular clusters and NGC 2808

| [Fe/H] | [α/Fe] | Y | $M_0$ (M$_\odot$) | $\Delta M$ (M$_\odot$) | $\tau_{postkink}$ ($10^6$ yr) | $\bar{\dot{M}}$ ($10^{-8}$ M$_\odot$ yr$^{-1}$) | $n_{dsh}$ |
|---|---|---|---|---|---|---|---|
| -1.5 | 0.4 | 0.250 | 0.81 | 0.216±0.022 | 35.01 | 0.617±0.063 | 12.370 |
| -1.5 | 0.4 | 0.258 | 0.81 | 0.246±0.028 | 33.53 | 0.734±0.084 | 12.328 |
| -1.5 | 0.4 | 0.275 | 0.81 | 0.285±0.024 | 30.23 | 0.943±0.079 | 12.230 |
| -1.1 | 0.4 | 0.250 | 0.83 | 0.110±0.024 | 47.58 | 0.231±0.050 | 12.923 |
| -1.1 | 0.4 | 0.252 | 0.83 | 0.126±0.022 | 46.70 | 0.270±0.047 | 12.910 |
| -1.1 | 0.4 | 0.290 | 0.77 | 0.203±0.023 | 40.54 | 0.501±0.057 | 12.729 |
| -1.1 | 0.4 | 0.340 | 0.71 | 0.231±0.022 | 32.70 | 0.706±0.067 | 12.456 |

A linear regression fit to our results, which are shown on fig. 3, gives the following result:

$$\frac{\bar{\dot{M}}}{10^{-8}\ \mathrm{M_\odot\ yr^{-1}}} = (11.30 \pm 1.34) - (0.854 \pm 0.106) n_{dsh}, \tag{4}$$

with an adjusted R square value of 0.915. Again, we find a strong (anti)correlation between the average post-kink mass loss rate and the number of convection zone density scale heights.

For this third sample of stars, the average mass loss rates actually overlap with the nominal rate mentioned in Section 5.

For all 15 entries in Tables 1-3, we find that the mean mass loss rate is $0.47 \times 10^{-8}$ M$_\odot$ yr$^{-1}$. Thus, overall, it appears that the nominal rate of mass loss predicted by MM19 is within a factor of about 2 of the average empirical mass loss rate. In view of the many uncertainties which entered into the MM19 prediction, it is remarkable that the average empirical mass loss rate turns out to differ from the predicted rate by such a small factor.

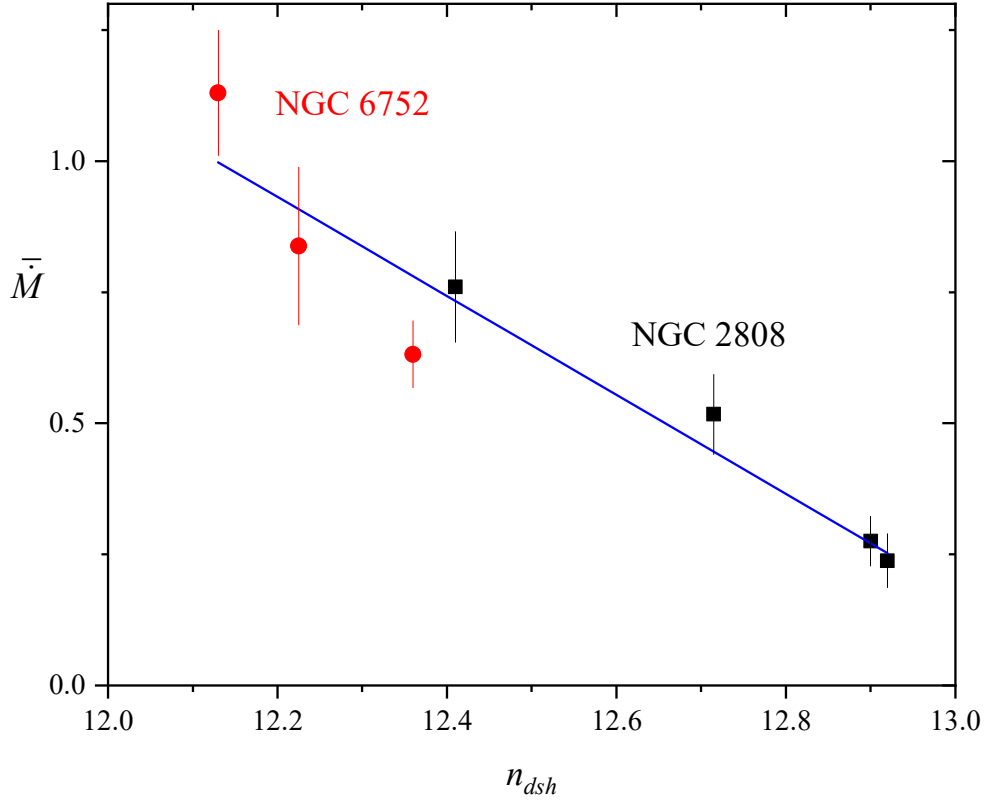

Figure 3. As fig. 1 but for stars in the multiple populations of the globular clusters NGC 2808 and NGC 6752. The blue line shows the best linear regression fit to the data.

*6.5. Dependencies of $\tau_{postkink}$ and $n_{dsh}$ on [Fe/H]*

Studies of the RGB bump in GC luminosity functions and theoretical modelling show that the kink occurs further up the RGB for clusters of lower metallicity (e.g. Cassisi & Salaris 1997). For models without enhanced helium abundance and with stellar masses relevant to this investigation, the post-kink RGB lifetime increases from ~20 Myr for [Fe/H] = - 2.4 to ~65 Myr for [Fe/H] = 0, and is approximately given by

$$\tau_{postkink} = 68.05 + 20.59[\text{Fe/H}] \text{ Myr.} \quad (5)$$

The number of density scale heights in the convection zone at the time of the kink also increases with [Fe/H] and is well approximated by

$$n_{dsh} = 14.360 + 1.551[\text{Fe/H}] + 0.146[\text{Fe/H}]^2. \quad (6)$$

Furthermore, $n_{dsh}$ decreases with evolution up the RGB and we find its value at the core flash is 0.86 ± 0.01 that at the kink, and the ratio is independent of [Fe/H].

## 7. Discussion and conclusions

A common approach to modelling mass loss on the RGB is to use the Reimers' mass formula (Reimers 1975) which is based on the idea that a fixed fraction of the radiation power goes into kinetic energy which drives mass loss from the stellar surface. If this hypothesis were correct, we would expect mass loss rates for a given stellar luminosity to be higher for stars with lower surface gravitational potential. Since, in general, stars of lower metallicity are more compact than higher metallicity counterparts, one might expect that (other things being equal) the integrated RGB mass loss $\Delta M$ should increase as metallicity increases. This is indeed the case for stars in globular clusters (e.g. Tailo et al. 2020). However recent results for field stars with **asteroseismic** determinations of integrated RGB mass loss find that $\Delta M$ *decreases* as metallicity increases (Brogaard et al. 2024; Li 2025).

Li (2025) has compared the averaged RGB mass loss rates predicted by the Reimers' formula and also from the formula of Schröder & Cuntz (2005) and the mass loss model of Cranmer & Saar (2011) with observed values, and finds that all of the theorical prescriptions fail to reproduce the observed metallicity dependence among field stars.

The current investigation is motivated by our earlier work on a model in which RGB mass loss is due to buoyant magnetic field loops produced by a shear dynamo operating below the convection zone (MM03b). Initially the flux loops are prevented from rising to the surface by the hydrogen discontinuity molecular gradient produced by the earlier inward excursion of the base of the convection zone. This molecular gradient barrier is removed when the hydrogen burning shell reaches the hydrogen discontinuity, allowing flux loops to rise to the surface. Not every loop will survive to reach the surface: diffusion and magnetic reconnection are expected to be continually operative as the loop traverses the outer layers of the convective envelope. We contend that the loops which do survive to reach the surface have the properties needed to initiate mass loss (see MM19). We have, therefore, looked for correlations between empirically inferred mass loss rates and properties of the convection zone at the time of the RGB kink (more precisely when the hydrogen discontinuity is removed). We find for three stellar samples, discussed in sub-sections 6.2 – 6.4, statistically significant anti-correlations between $\overline{\dot{M}}$, the average RGB mass loss rate (assuming it begins at the kink) and $n_{dsh}$, the number of density scale heights in the convection zone at the time of the kink. This result is highly suggestive of a strong dependence of the mass loss rate on the number of density scale heights in the convection zone for the post-kink RGB evolution. We propose that it is the number of density scale heights in the convection zone that controls the mass loss rate starting after the RGB kink.

In such a scenario, the integrated RGB mass loss is determined by two factors: 1) The post-kink RGB lifetime, and 2) the mass loss rate which depends on the number of convection zone density scale heights. We can now qualitatively understand the trend of integrated mass loss with [Fe/H]. **At very low** [Fe/H], $n_{dsh}$ is relatively small and mass loss rates are large. However, the short post kink lifetime leads to a small value for the integrated mass loss. As [Fe/H] increases, $\tau_{postkink}$ also increases. The mass loss rate will decrease but not enough to prevent $\Delta M$ from increasing. Near [Fe/H] ~ -0.8, the mass loss rate starts to decrease more quickly with increasing [Fe/H] (because of the nonlinear dependence of $n_{dsh}$ on [Fe/H]), so that, even though $\tau_{postkink}$ is increasing with [Fe/H], $\Delta M$ begins to decrease.

Quantitatively, we find that, for three samples of stars, the mean mass loss rate of stars on the RGB above the "bump" agree within a factor of about 2 with the value which was predicted by MM19 for a plasmoid-driven wind.

We find that the statistical significance of the anticorrelation between $\bar{\dot{M}}$ and $n_{dsh}$ is weaker for the first-generation globular cluster stars than for the other two stellar samples. As can be seen in fig. 2, there is an indication that there is a turnover in the $\bar{\dot{M}}$ and $n_{dsh}$ relation at the two lowest $n_{dsh}$ values, which correspond to the two lowest values of [Fe/H]. The reason for this turnover is not clear but we note that there is no well-defined kink in the HRD on the RGB for these low [Fe/H] values and, also, the values of $\tau_{postkink}$ at low [Fe/H] are longer than would be from extrapolation of values at higher [Fe/H]. Why this is the case needs further study.

We have not included convective overshooting in our stellar modelling. A thorough study of the effects of convective overshooting in our mass loss model is beyond the scope of this paper, but we note that recent work (Briganti et al. 2025) has explored how the location in the HRD of the RGB bump depends on the efficiency of mixing due to convective overshooting. Briganti et al. find that for more efficient overshooting the discontinuity in the hydrogen mass fraction profile occurs deeper in the stellar envelope and results in a less luminous RGB bump. A less luminous RGB bump results in an increase in the time for the star to evolve from the kink to the helium core flash, which would decrease the average post-kink mass loss rates. Provided convective overshooting reduces the RGB bump luminosity in a systematic way, as seems to be the case in the findings of Briganti et al., our conclusions will not be qualitatively changed.

## Acknowledgements

JM acknowledges partial support from the NASA Delaware Space Grant Consortium. We thank the anonymous referee for comments that led to improvement of the manuscript.

## ORCID iDs

J. MacDonald https://orcid.org/0000-0002-0506-5124
D. J. Mullan https://orcid.org/0000-0002-7087-9167